\documentclass{emulateapj}
\usepackage{epsfig, natbib}
\newcommand{\tdn}{\tau_{\rm{dyn}}}
\newcommand{\tdf}{\tau_{\rm merge}}

\newcommand{\vorb}{v_{\rm{orb}}}

\newcommand{\msat}{M_{\rm{sat}}}
\newcommand{\mhost}{M_{\rm{host}}}
\newcommand{\mstar}{M_{\star}}

\newcommand{\rvir}{r_{\rm{vir}}}

\newcommand{\re}{R_e}

\newcommand{\logl}{\ln \Lambda}

\newcommand{\dd}{{\rm d}}

\shorttitle{Dynamical Friction}
\shortauthors{Boylan-Kolchin, Ma, \& Quataert}
\slugcomment{Draft version -- \today}
\begin{document}

\title{Dynamical friction and galaxy merging timescales}

\author{Michael Boylan-Kolchin\altaffilmark{1}, Chung-Pei
  Ma\altaffilmark{2}, and Eliot Quataert\altaffilmark{3}}
\affil{Department of Astronomy, University of California,
  Berkeley, CA 94720}
\altaffiltext{1}{mrbk@berkeley.edu}
\altaffiltext{2}{cpma@berkeley.edu} 
\altaffiltext{3}{eliot@astro.berkeley.edu} 

\begin{abstract}
  The timescale for galaxies within merging dark matter halos to merge with each
  other is an important ingredient in galaxy formation models.  Accurate
  estimates of merging timescales are required for predictions of astrophysical
  quantities such as black hole binary merger rates, the build-up of stellar
  mass in central galaxies, and the statistical properties of satellite galaxies
  within dark matter halos.  In this paper, we study the merging timescales of
  extended dark matter halos using $N$-body simulations.  We compare these
  results to standard estimates based on the Chandrasekhar theory of dynamical
  friction.  We find that these standard predictions for merging timescales,
  which are often used in semi-analytic galaxy formation models, are
  systematically shorter than those found in simulations.  The discrepancy is
  approximately a factor of 1.7 for $\msat/\mhost \approx 0.1$ and becomes
  larger for more disparate satellite-to-host mass ratios, reaching a factor of
  $\sim 3.3$ for $\msat/\mhost\approx 0.01$.  Based on our simulations, we
  propose a new, easily implementable fitting formula that accurately predicts
  the timescale for an extended satellite to sink from the virial radius of a
  host halo down to the halo's center for a wide range of $\msat/\mhost$ and
  orbits.  Including a central bulge in each galaxy changes the merging
  timescale by $\la 10\%$.  To highlight one concrete application of our
  results, we show that merging timescales often used in the literature
  overestimate the growth of stellar mass by satellite accretion by $\approx 40
  \%$, with the extra mass gained in low mass ratio mergers.
\end{abstract}


\section{Introduction}

As originally formulated by \citet{chandrasekhar1943}, the deceleration of
an orbiting point mass ```satellite,'' due to dynamical friction on a
uniform background mass distribution, is given by
\begin{equation}
  \label{eq:df}
  \frac{d}{dt}\vec{v}_{\rm{orb}} = -4 \pi G^2 \ln (\Lambda) \, \msat
  \, \rho_{\rm host}(<\vorb) \, \frac{\vec{v}_{\rm{orb}}}{\vorb^3} \,,
\end{equation}
where $\rho_{\rm host}(<\vorb)$ is the density of background particles with
velocities less than the orbital velocity $v_{\rm orb}$ of the satellite,
$\msat$ is the mass of the satellite, and $\Lambda$ is the usual Coulomb
logarithm (e.g, \citealt{chandrasekhar1943, white1976}).

Accurate estimates of the effects of dynamical friction and the timescale
for an orbiting satellite to lose its energy and angular momentum to merge
with a host are essential for many astrophysical problems.  A thorough
understanding of dynamical friction is particularly critical for
semi-analytic models (SAMs) of galaxy formation.  The growth of dark matter
halos via mergers can be computed either analytically or via
dissipationless cosmological simulations, but the growth of galaxies
depends on their dynamical evolution within larger dark matter halos.  As a
result, dynamical friction (and tidal stripping, which determines $\msat$)
provides a critical link between dark matter halo mergers and the galaxy
mergers that determine, e.g., stellar masses, supermassive black hole
masses, galaxy colors, and galaxy morphologies.  Since it is
computationally infeasible to simulate both dark matter on cosmological
scales and baryonic physics relevant to galaxies, SAMs will remain an
important tool for interpreting observations of galaxy formation and
evolution for the foreseeable future.

There are, however, often significant uncertainties in directly applying
equation (\ref{eq:df}) due to the approximate nature of the Chandrasekhar
formula, and ambiguities in the appropriate value of the Coulomb logarithm
and the definition of the satellite mass $\msat$ for extended objects
(e.g., galaxies or dark matter halos).  SAMs, for instance, typically use
mild variations on the same basic formula for modeling the merger timescale
$\tdf$ induced by dynamical friction:
\begin{equation}
  \label{eq:dfSAM}
  \frac{\tdf}{\tau_{\rm dyn}} =1.17 \, \frac{f_{\rm df}\, \Theta_{\rm orb}}
  {\logl} \frac{\mhost}{\msat} \,  
\end{equation}
(e.g., eq.~7.26 of \citealt{binney1987}), where  
$\Theta_{\rm orb}$ contains information about the orbital energy and
angular momentum, $f_{\rm df}$ is an adjustable parameter, and $\tau_{\rm
  dyn}$ is the dynamical timescale at the virial radius $\rvir$ of the host
halo, related to the circular velocity at $\rvir$, $V_c(\rvir)$, by
\begin{equation}
  \label{eq:taudyn}
   \tau_{\rm dyn}\equiv \frac{\rvir}{V_c(\rvir)}
           =\left( \frac{\rvir^3}{G\mhost} \right)^{1/2} \,.
\end{equation}
Differences among different SAMs enter mainly in how each model treats $\logl$,
$\Theta_{\rm orb}$, and $f_{\rm df}$, as well as how $\msat$ is determined and
when $\tdf$ is determined (e.g., \citealt{kauffmann1993, somerville1999,
  cole2000, croton2006}).  Uncertainties in these parameters are reflected in
equally large uncertainties in $\tdf$.  

The Coulomb logarithm, for example, should technically be expressed as ${1\over
  2} \, \ln(1+\Lambda^2)$; using $\ln \Lambda$ is appropriate in the limit of
large $\Lambda$ \citep{binney1987}.  The Coulomb logarithm represents the ratio
of the largest and smallest impact parameters of field stars that contribute to
small-angle scatterings of the satellite: $\Lambda=b_{\rm max}/b_{\rm min}$.
This motivates a conventional choice for the Coulomb logarithm:
$\Lambda=1+\mhost/\msat$,
where $\msat$ is often taken to be the mass of
the satellite when it enters the virial radius of the host halo of mass $\mhost$.
Alternatively, the Coulomb logarithm can be taken to be a
mass-independent constant (e.g., \citealt{cooray2005}).  When using equation
(\ref{eq:df}) to model the full orbits of satellites, another approach is to
assume that the Coulomb logarithm varies over the course of the orbit (e.g.,
\citealt{hashimoto2003,zentner2005a,fellhauer2007}). These various prescriptions
for the Coulomb logarithm can differ by a factor of 2-3.  

In addition, it is known that $\msat$ cannot refer to just the bound mass for
extended objects because tidally stripped material still in the vicinity of the
satellite also contributes to dynamical friction
\citep{fujii2006,fellhauer2007}.  The
uncertainties in $\tdf$ introduced by different prescriptions for $\logl$ and
$\msat$ are important for galaxy formation models because $50\%$ changes in the
timescale for galaxies to merge within dark matter halos could significantly
change the predictions for galaxy growth via merging or cannibalism, black hole
merger rates, and the evolution of satellite galaxies in groups and clusters.

Full dynamical models of the evolution of merging halos -- complete with the
physics of dynamical friction, tidal stripping, and gravitational shocking --
have been created by a number of groups (e.g., \citealt{taylor2001, taffoni2003,
  zentner2005a, gnedin2003}).  These models, however, are based on approximate
treatments of the underlying physical processes, and obtaining merging
timescales would often require numerically integrating individual satellite
orbits \citep{velazquez1999}.  In this paper we take a different approach and
instead compute merger timescales directly using numerical simulations.  We
focus on the range of satellite masses and orbital parameters that are relevant
for dark matter halo mergers in hierarchical galaxy formation models.  Our goal
is to provide a fitting formula that is as simple as possible but that still
accurately reflects the physics of dynamical friction and tidal stripping.

The remainder of this paper is organized as follows.  Section 2 describes the
numerical simulations we have performed.  In Section 3, we investigate the
merging timescales $\tdf$ computed from our numerical simulations, propose a new
fitting formula for $\tdf$, and compare to previous results.  Section 4 contains
two sample applications of our proposed fitting formula: an estimate of the mass
spectrum and orbital distribution of merging galaxies.  Our results and
conclusions, and their implications, are reviewed in Section~5.  Throughout this
paper, we use ``virial'' quantities that are defined relative to $200 \rho_c$,
where $\rho_c$ is the critical density.  When necessary, we assume a cosmology
of $\Omega_m=0.3$, $\Omega_{\Lambda}=1-\Omega_m=0.7$, and $H_0=70 \,{\rm km \,
  s^{-1}\, Mpc^{-1}}$.

\section{Satellite decay from numerical simulations}
\label{sec:sims}
\subsection{Description of simulations and initial conditions}
We have performed a number of numerical simulations to test the agreement
between merger timescales predicted by equation~(\ref{eq:dfSAM}) and those
derived directly from $N$-body simulations.  Each of our simulations consists of
a host \citet{hernquist1990} halo and a satellite Hernquist halo; the ratio of
satellite to host mass and the initial orbital parameters of the satellite are
varied from run to run.  The host and satellite halos were constructed using
$N_{\rm host}=2\times 10^5$ particles for the host and equal particle masses for
the host and satellite particles, i.e., $N_{\rm sat}=(\msat/\mhost) \, 2 \times
10^5$.  Each halo was constructed by sampling the full phase-space distribution
function (under the assumptions of spherical symmetry and velocity isotropy) and
was tested to be stable when evolved in isolation.  In addition, two simulations included a (self-consistently constructed) Hernquist bulge in both the host and satellite to test the effects of baryonic components on merging timescales.

The numerical simulations were run using {\sc gadget-2} \citep{springel2005}.
Outputs were saved every 0.1 Gyr, or equivalently, $\approx 0.06 \, \tdn$.  To
ensure that our simulations are not affected by numerical artifacts, we have
performed several convergence runs with respect to both the particle number (up
to $10^6$ particles in the host halo) and the force softening $\epsilon$ (which
is reduced by up to a factor of 5 from our fiducial value of $\epsilon=10^{-3}\,
r_{\rm vir}$).  Because our simulations involve only gravity, the length, mass,
and time scales can be rescaled for any host halo mass; for definiteness,
however, we quote our results for host halos of total mass $\mhost =10^{12}\,
M_{\odot}$ and Hernquist scale length $a=40$ kpc.  Matching to a
\citet[hereafter NFW]{navarro1997} profile with a virial
mass equal to the Hernquist halo's total mass and an identical
density at $r \ll a$, our standard host halo corresponds to an NFW halo with a
concentration $\approx 8.5$ (see, e.g., \citealt{springel2005a} for more details
on how to relate Hernquist and NFW halos).  A comparison run using an NFW halo
had a merging timescale within 5\% of that from the corresponding run with a
Hernquist halo.

\begin{deluxetable}{ccccc}
\tablecaption{List of Simulations
\label{table:ICs}}
\tablehead{
\colhead{{\bf Run}}
& \colhead{${\bf \msat/\mhost}$}
& \colhead{{\bf circularity} \boldmath{$\eta$}}
& \colhead{\boldmath{$r_c(E)/\rvir$}}
& \colhead{\boldmath{$\tdf$}}\\
(1) & (2) & (3) & (4) & (5)
}
\startdata
3b   &  0.3       &     0.78  &    1.0  &  4.4 \\
3d   &  0.3       &     1.00  &    1.0  &  6.9 \\
5b   &	0.2	  & 	0.46  &    1.0  &  3.5 \\
5c   &	0.2	  &	0.65  &    1.0  &  4.5 \\
5d   &	0.2	  &      0.78 &    1.0  &  6.0 \\
5e   &	0.2	  &      1.00  &    1.0  &  9.5 \\	
10a  &   0.1	  &      0.33 &    1.0  &  4.85 \\	
10a$\star$ &   0.1 	  &      0.33 &    1.0  &  4.3 \\
10b  &   0.1	  &      0.46 &    1.0  &  6.0 \\	
10c  &   0.1	  &      0.65 &    1.0  &  8.25 \\
10c$\star$ &   0.1	  &      0.65 &    1.0  &  7.8 \\
10d  &   0.1	  &      0.78 &    1.0  &  10.75 \\     
14b  &   0.0707   &      0.46 &    1.0  &  9.0 \\
14c  &   0.0707   &      0.65 &    1.0  &  12.5 \\
20a  &   0.05	  &      0.33 &    1.0  &  10.0 \\
20b  &   0.05	  &      0.46 &    1.0  &  11.0 \\
20c  &   0.05	  &      0.65 &    1.0  &  17.0 \\
40b  &   0.025	  &      0.46 &    1.0  &  22.0 \\
40c  &   0.025    &      0.65 &    1.0  &  38.0 \\
10a1 &   0.1      &      0.33 &    0.65 &  3.5 \\
10a4 &   0.1      &      0.33 &    0.80 &  4.3\\
10b1 &   0.1      &      0.46 &    0.65 &  4.3 \\
10b4 &   0.1      &      0.46 &    0.80 &  5.0 \\
10c1 &   0.1	  &      0.65 &    0.65 &  5.6 \\  
10c2 &   0.1	  &      0.65 &    0.70  &  5.5 \\   
10c3 &   0.1	  &      0.65 &    0.75 &  6.2 \\   
10c4 &   0.1	  &      0.65 &    0.80  &  6.35 \\ 
10c5 &   0.1	  &      0.65 &    0.90  &  7.5 \\ 
\enddata
\tablecomments{
Description of columns:\\
(1) Name of simulation\\
(2) Ratio of initial satellite mass to initial host halo mass\\
(3) Initial orbital angular momentum, parametrized by circularity $\eta$\\
(4) Initial orbital energy, parametrized by $r_c(E)/\rvir$ \\
(5) Dynamical friction merging time $\tdf$, in Gyr, for a host with virial
mass $\mhost=10^{12} \,M_{\odot}$, measured from the simulation as
described in Sec.~\ref{sec:sims}\\
($\star$ indicates run with baryonic component; see \S\ref{subsec:baryons} for
details)}
\end{deluxetable}

In the absence of dynamical friction, the orbit of a test particle in a
static spherical potential is entirely specified by its energy $E$ and
angular momentum $J$.  An equivalent parametrization for bound orbits is to
use the orbital circularity 
\begin{equation}
   \eta \equiv \frac{j}{j_c(E)} \,,
\end{equation}
which is the specific angular momentum relative to the specific angular
momentum of a circular orbit with the same energy [$\eta$ is related to
eccentricity $e$ by $\eta=(1-e^2)^{1/2}$], and $r_c(E)$, which is the radius of
a circular orbit with the same energy as the particle in question.  Since
dynamical friction dissipates energy, the orbit of a satellite galaxy in a host
galaxy generally depends on its initial position $r(t=0)$ in addition to
$E(t=0)$ and $J(t=0)$.

We explore a range of orbital circularity $\eta$, energy $r_c(E)$, and
satellite-to-host mass ratios $\msat/\mhost$ in our simulations.  A summary
of the production runs and the resulting $\tdf$ is presented in
Table~\ref{table:ICs}.  The ranges of the parameters are chosen so that the
satellites can plausibly undergo significant orbital evolution within a
Hubble time (see Sec. \ref{sec:range} for more details).  Systems with mass
ratios $\msat/\mhost$ much smaller than 1:40 are therefore dynamically
uninteresting and excluded.

\subsection{Defining timescales: angular momentum loss}
The timescale for a satellite to merge with its host halo can be defined in
a number of different ways.  One common definition is to take a fiducial
radius of the baryonic component assumed to reside at the center of the
halo and assume that the satellite has merged when its separation from the
host's center equals this radius.  In this work, we instead consider a
satellite merged when it has lost all of its specific angular momentum $j
\equiv r\, v_t$ (relative to the host).  In practice, this definition
agrees with the more commonly-used one in most cases but also works well
for situations in which the definition based on orbital separation can give
undesired results.  Consider highly eccentric orbits, in which the
satellite can come very close to the center of the host while retaining
significant orbital energy; satellites on these orbits often do not merge
for multiple dynamical times following the first close encounter.  The
specific angular momentum of the satellite should, however, be a
non-increasing function of time, so using $j(t)/j_0$ [where $j_0=j(t=0)$]
is therefore our preferred definition for merging.

\begin{figure}
  \centering
  \includegraphics[scale=0.55, bb=70 156 516 612]{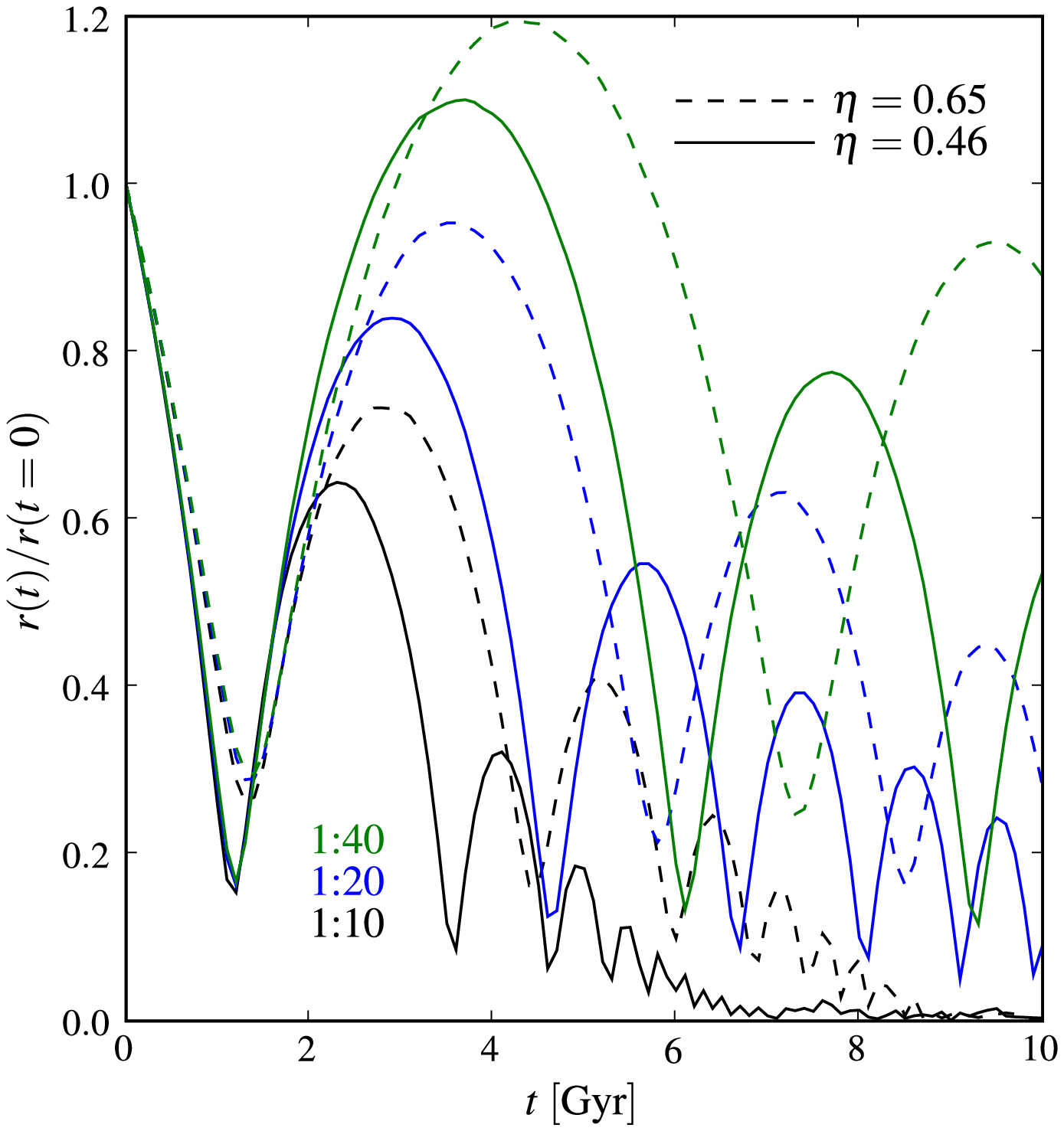}
  \includegraphics[scale=0.55, bb=70 175 516 595]{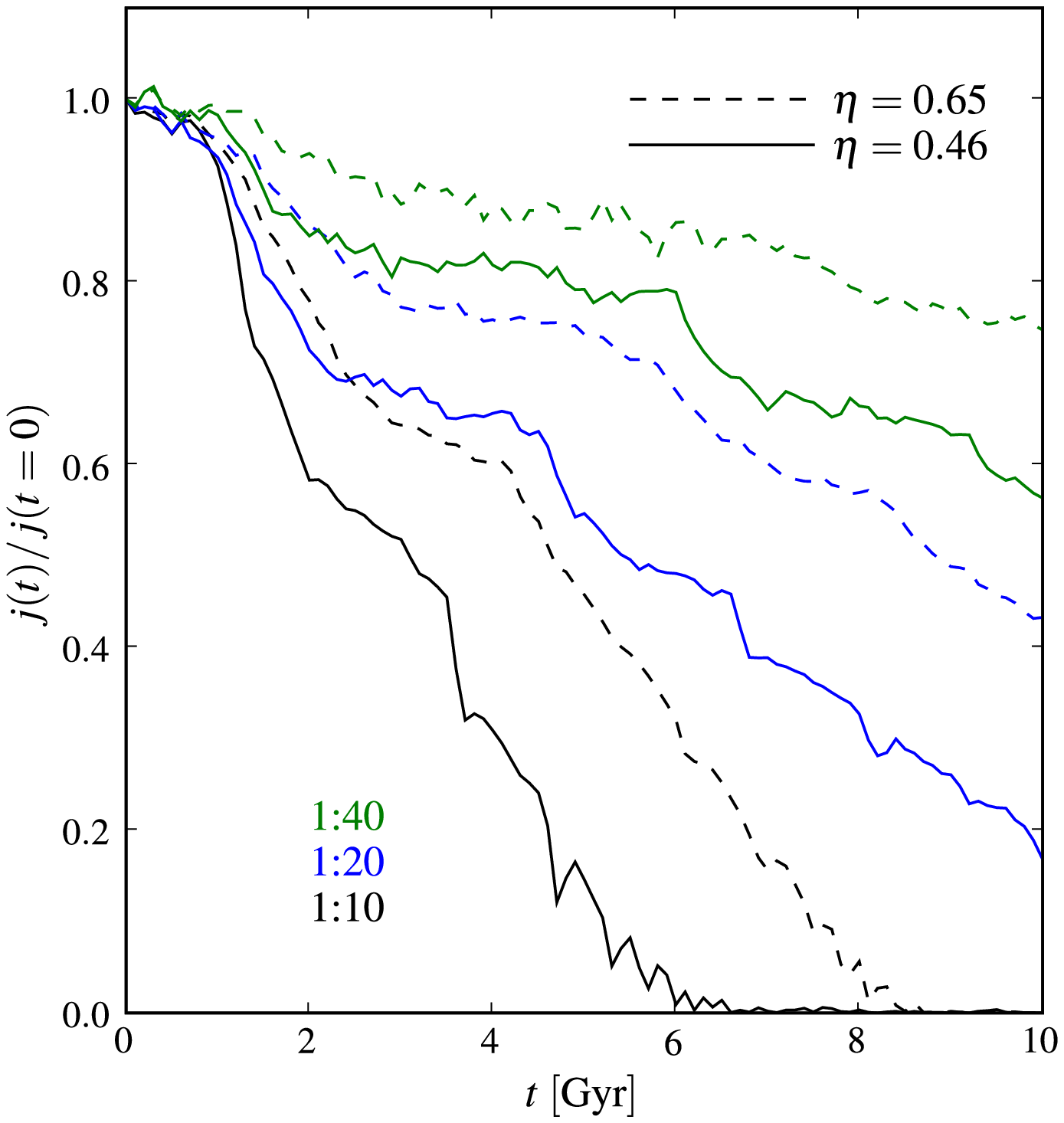}
  \caption{Top: Trajectories of satellites (i.e., separation between
    satellite and host centers) from six simulations with two different
    orbits -- $\eta=0.46$ (solid) and 0.65 (dashed) -- and three different
    mass ratios: 1:10 (black), 1:20 (blue), and 1:40 (green).  For all
    cases, satellites on more eccentric orbits (lower $\eta$) or with
    higher mass ratios merge more quickly.  Bottom: angular momentum decay
    for the same satellites. The loss of angular momentum initially
    correlates well with pericentric passages but at later times $dj/dt
    \sim \, {\rm constant}$.\label{fig:ang_all}}
\end{figure}

Fig.~\ref{fig:ang_all} shows the trajectories (top) and angular momentum
decay (bottom) for three sets of live-satellite simulations starting from
$\rvir$, with $\msat/\mhost=0.025,\, 0.05, \, {\rm and} \; 0.10$, for two
different orbital circularities, $\eta=0.46$ (solid curves) and $\eta=0.65$
(dashed curves).  For a given circularity, systems with more disparate
masses merge much more slowly, as is expected from basic dynamical friction
considerations.  For a given mass ratio, the satellite on the more
eccentric orbit (i.e. smaller $\eta$) loses angular momentum faster,
resulting in a more rapid merger.  The
difference is non-negligible even for mass ratios as similar as 1:3.  In this
case, a satellite with $\eta=0.78$ merges in approximately 4.4 Gyr while a
satellite on a circular orbit takes over 50\% longer -- 6.9 Gyr --
to merge (see last column of Table~1).

Fig.~\ref{fig:ang_all} highlights the angular momentum loss process.
A comparison between the top and bottom panels shows that the bulk of the
angular momentum loss initially coincides with pericentric passages (e.g.,
at 1 and 4.5 Gyr for the 1:20 blue curves) and the accompanying tidal
stripping and shocking (see also \citealt{boylan-kolchin2007}).  It is also
interesting to note that while the angular momentum loss is initially
somewhat impulsive, it later ($t \ga 5$ Gyr) becomes nearly constant in
time.
Physically, the approximate constancy of $dj/dt \sim j/\tdf(r)$ follows
from the fact that the merger timescale at any radius is roughly linearly
proportional to radius (see eq. [\ref{eq:tdf_fit}] discussed below), as is
$j$ itself.  It is also interesting to note that a satellite can continue
to orbit for several giga-years even after its initial orbit has
dipped within 5\% of $\rvir$ of the host (e.g., the 1:20 merger in
Fig.~\ref{fig:ang_all}).  In fact, at the first pericentric pass within
$\sim 5\%$ of $\rvir$ (which occurs at $t \approx 6.5$ Gyr for $\eta =
0.46$), the satellite has only lost 60\% of its initial angular momentum.
Such satellites should \emph{not} be considered merged even though they may
pass very close to the center of the halo (within the radius at which a
central galaxy might lie); our calculations show that these satellites will
orbit for a significantly longer time before actually merging.

When the mass of the satellite is much less than the mass of the host, $\tdf$
becomes prohibitively long to study with simulations, as numerical relaxation
becomes an important effect.  For these cases (only three runs: 20c, 40b, and
40c), we determine $\tdf$ by linearly extrapolating $j(t)/j_0$.  This tends to
be a reasonable measure of the merging time as long as we extrapolate
\emph{after} the first two pericentric passages, where gravitational shocks and
tidal stripping cause substantial changes in the angular momentum.  After the
first two pericentric passages, the angular momentum loss occurs at a roughly
uniform rate.

\subsection{Effects of baryons}
\label{subsec:baryons}
\begin{figure}
  \centering
  \includegraphics[scale=0.55]{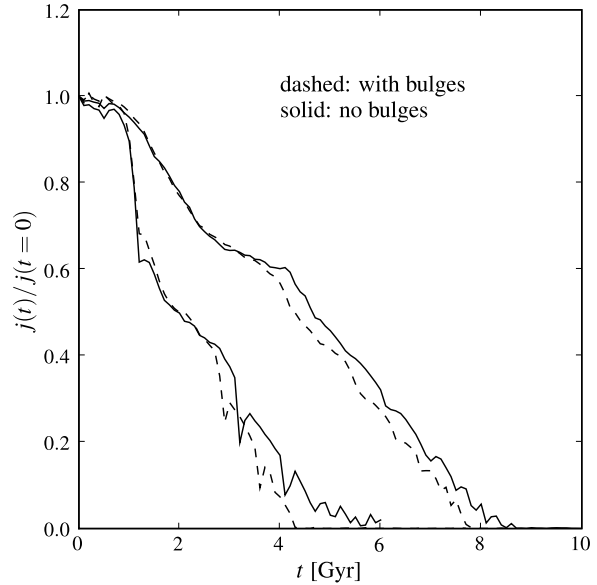}
  \caption{Effects of including baryonic components.  The solid curves show the
    angular momentum loss for runs 10a (left; $\eta=0.33$) and 10c (right;
    $\eta=0.65$).  The dashed curves show $j(t)/j_0$ for runs that have a
    stellar bulge (with $M_{\rm vir}/M_{\rm bulge}=20$) in both the host and
    satellite but are otherwise identical.  Including baryons leads to merger
    timescales that are $\la 10\%$ shorter than those of the corresponding dark
    matter-only simulations.}
  \label{fig:stars}
\end{figure}
Although the ratio of a galaxy's stellar mass to the virial mass of its dark
matter halo does not exceed $\approx 0.1$, it is conceivable that neglecting
baryons could lead to significant errors in the predicted merging timescales.
In particular,
bulges in elliptical galaxies are significantly more dense than dark matter
halos at the same physical scale, meaning that bulges are more resistant to
disruption via tidal shocking.  Thus, the expectation is that merger
timescales would be shorter for full galaxy models than for dark matter halos
alone.  

In order to test the effects of the baryonic components of galaxies on merging
timescales, we ran two additional simulations.  Both simulations used
$\msat/\mhost=0.1$, but we included stellar bulges with $M_{\star}/M_{\rm
  dm}=0.05$ in both the host and satellite for each simulation.\footnote{The
  bulge + dark matter halo systems for the initial conditions are set up
  self-consistently and are very stable over many dynamical times when evolved
  in isolation; see \citet{boylan-kolchin2007} for details.}  The effective
radii of the bulges were chosen to be representative of those measured by
\citet{shen2003} from the Sloan Digital Sky Survey \citep{york2000}, $\re
\propto \mstar^{0.56}$, and the orbits of the two runs are identical to runs 10a
and 10c.

In Fig.~\ref{fig:stars}, we compare the angular momentum loss as a function of
time for the runs without stars (solid curves) and with stars (dashed
curves). Including stars does lead to faster mergers, as expected, but the
difference in merging timescales is less than 10\%, which is quite modest. When
the satellite first enters the host's virial radius, the bulges contribute
negligibly to dynamical friction drag because the baryons contribute $< 10\%$ to
the system's total mass. At late stages in the merger, the bulges may be more
important, as the dark matter halo is more efficiently stripped than the
bulge. The satellites spend the vast majority of their time at large radii,
however, so that the error introduced by ignoring baryons is small when
considering the total time required to merge from the virial radius (see
Fig.~\ref{fig:stars}).


\section{Computing merger timescales}

In this section we provide a simple parameterization of the merging
timescales determined from the numerical simulations discussed above.
Comparisons between this new fitting formula and some commonly used ones in
the literature are also presented (Sec~3.3).

\subsection{A fitting formula}

We fit the merger timescales from the simulations in Table~1 to a simple
formula,
\begin{equation}
   \frac{\tdf}{\tau_{\rm dyn}} = A \, {(\mhost/\msat)^b \over \ln(1+\mhost/\msat)}
  \exp\left[c \, {j \over j_c(E)} \right] \, \left[{r_c(E) \over \rvir} \right]^d
  \label{eq:tdf_fit} \;,
\end{equation}
where the constants $b, \, c$, and $d$ parameterize the dependence of the
merger timescales on the host-to-satellite mass ratio $\mhost/\msat$,
orbital circularity $\eta = j/j_c(E)$, and orbital energy $r_c(E)$,
respectively.  Both $\eta$ and $r_c(E)$ are computed self-consistently
using the Hernquist potential, {\it not} using the two-body
approximation. To avoid ambiguities in the definition of satellite and host
mass as a function of the location of the satellite (or time), we define
the masses $\mhost$ and $\msat$ here as the {\it virial} masses of the host
and the satellite when entering the host's virial radius.  For the same
reason, we use the specific angular momentum $j$ rather than the full
angular momentum $J$ in computing the orbital properties of the satellites;
since the circularity depends only on a ratio of angular momenta, this
choice does not affect calculations of $\eta$.  The dynamical time $\tdn$
is the dynamical time at the host's virial radius $\rvir$ given by
equation~(\ref{eq:taudyn}).  Note that $\tdn=0.1 \, H^{-1}$ (where $H$ is
the Hubble constant) independent of the host halo mass; at $z=0$, $\tdn=1.4$
Gyr for $H=70 \,{\rm km \, s^{-1}\, Mpc^{-1}}$.  

With the definitions above, we find that
\begin{equation}
  \label{eq:abcd}
  A=0.216\,, \quad b=1.3\,, \quad c=1.9\,,\quad d=1.0
\end{equation}
provide fits to the simulation results with a standard deviation of $< 7\%$
and a maximum deviation of 12.5\%, a significant improvement over
equation~(\ref{eq:dfSAM}).  To obtain the values in
equation~(\ref{eq:abcd}), we have chosen to fit to the parameter $d$
separately from $A, \, b,$ and $c$.  For $A$, $b$, and $c$, we have fitted
to $\tdf$ from the subset of simulations in which the satellites begin at
the virial radius of their host with an orbital energy $E=E_c(\rvir)$,
i.e., with $r_c(E)=\rvir$, and the fit is independent of $d$.  To determine
the value of $d$ in equation~(\ref{eq:abcd}), we use the series of
simulations with $M_{\rm sat}/M_{\rm host}=0.1$ and
$r_c(E)/\rvir=[0.65,\,0.7,\,0.75,\, 0.8, \, 0.9, \, 1.0]$ (all starting at
$\rvir$).  Holding $A$, $b$, and $c$ fixed, we investigate whether any
single choice of $d$ results in a consistently accurate prediction of the
merging timescale.

\begin{figure}
  \centering
  \includegraphics[scale=0.55]{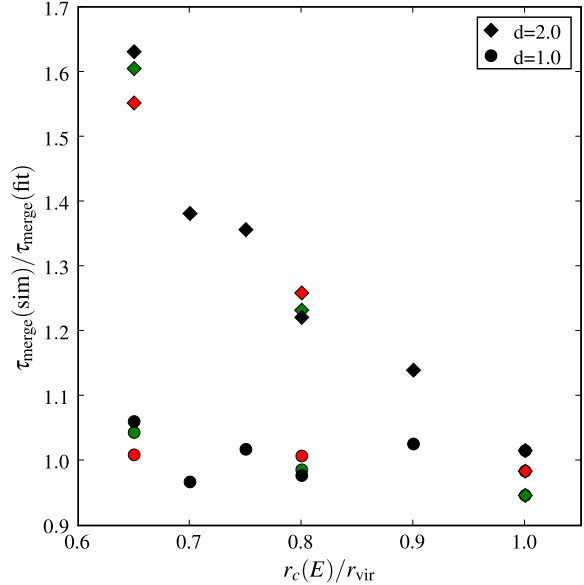}
  \caption{Comparison of merger time measured in our $N$-body simulations, $\tdf
    ({\rm sim})$, with the predicted $\tdf ({\rm fit})$ based on
    eq.~(\ref{eq:tdf_fit}), as a function of orbital energy $r_c(E)/\rvir$.  The
    orbital circularity is color-coded: $\eta=0.33$ (red), 0.46 (green), and
    0.65 (black).  
    The circles use the best-fit parameters in
    eq.~(\ref{eq:abcd}) and show the close agreement in $\tdf$ between the
    simulations and the fits.  The diamonds, on the other hand, illustrate the
    poor match when the energy dependence in eq.~(\ref{eq:tdf_fit}) is taken to
    be the conventional $d=2.0$ instead of $d= 1.0$.  All points are for
    $\msat/\mhost=0.1$.}
  \label{fig:dftimes_pred}
\end{figure}

\begin{figure}
\centering
\includegraphics[scale=0.55]{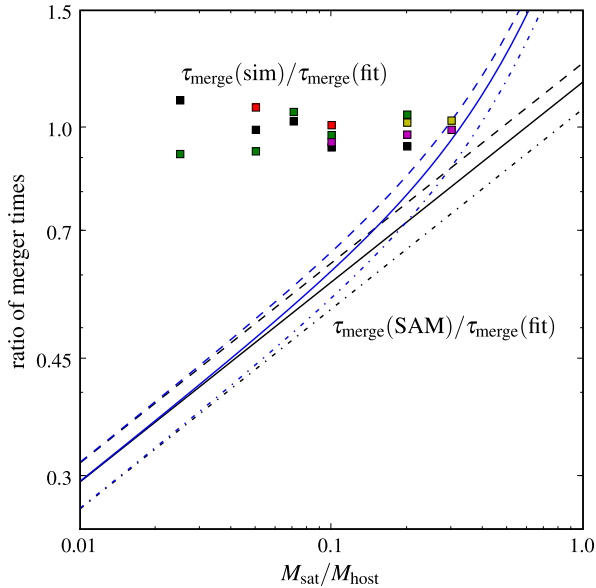}
\caption{Comparison of merger times computed from $N$-body simulations, fitting
  formula eq.~(\ref{eq:tdf_fit}), and a fiducial SAM using eq.~(\ref{eq:dfSAM})
  and two choices of Coulomb logarithm (black and blue curves), for a variety
  of satellite-to-host mass ratios $\msat/\mhost$ and orbital circularities
  $\eta$, with $r_c(E)/\rvir=1$.  The square symbols show the ratio $\tdf ({\rm
    sim})/\tdf ({\rm fit})$ and illustrate that the simulations and fit agree
  well, with an error of $\la 10\%$.  The orbital circularities are color-coded:
  $\eta=0.33$ (red), 0.46 (green), 0.65 (black), 0.78 (magenta), and 1.0
  (yellow).  The fiducial SAM, on the other hand, generally underestimates the
  merging time, particularly at small $\msat/\mhost$, as shown by the three
  curves for different circularities: $\eta=0.25$ (solid), 0.5 (dashed), and
  0.75 (dot-dashed).  The SAM here assumes $f_{df}=1$, $\Theta_{\rm
    orb}=\eta^{0.78}$, and $r_c(E)/\rvir=1$.  The different color curves
  correspond to different choices of Coulomb logarithms for the SAM: $\ln
  (1+\mhost/\msat)$ [black curves] and ${1 \over 2} \ln (1+\mhost^2/\msat^2)$
  [blue curves]}
  \label{fig:sim_sam_fit}
\end{figure}

The close agreement between the numerical simulations and the fits in
equation~(\ref{eq:tdf_fit}) is illustrated by the circles in
Fig.~\ref{fig:dftimes_pred} and squares in Fig.~\ref{fig:sim_sam_fit}.  The two
figures show the ratio of the timescale measured from each simulation, $\tdf
({\rm sim})$, to the timescale predicted by equation~(\ref{eq:tdf_fit}), $\tdf
({\rm fit})$, for a wide range of orbital energy ($0.65\le r_c(E)/\rvir \le
1.0$), orbital circularity ($0.33 \le \eta \le 1$), and satellite-to-host halo
mass ($0.025 \le \msat/\mhost \le 0.3$).  In addition,
Fig.~\ref{fig:dftimes_pred} shows that using $d=1.0$ (circle symbols) in
equation~(\ref{eq:tdf_fit}) results in good agreement between predicted and
measured $\tdf$, while the often used $d=2.0$ (diamond symbols) results in a
poor match, systematically underestimating $\tdf$ (see \S
\ref{subsubsec:c_and_d} for more discussion).  In Fig.~\ref{fig:sim_sam_fit}, we
also compare $\tdf ({\rm fit})$ to that predicted by a fiducial SAM, $\tdf ({\rm
  SAM})$; this is discussed in \S \ref{sec:compare} below.

A further point of interest is that using $d \approx 1$ in
Equation~(\ref{eq:tdf_fit}) also provides a significantly better fit than $d=2$
when considering the merging timescale as a function of position along a given
orbit.  This suggests that equation~(\ref{eq:tdf_fit}) can be used for starting
radii other than $\rvir$, though its accuracy will certainly decline for $r \ll
\rvir$.  This possibility is not explored in any more detail in this paper,
however, and we restrict ourselves to starting radii of $\rvir$.

\subsection{Range of validity}

\label{sec:range}

We have chosen to simulate parameters that correspond to probable satellite
orbits in a $\Lambda$CDM cosmology.  The mass ratios we have considered --
$0.025 \le \msat/\mhost \le 0.3$ -- cover a reasonable range, as lower mass
ratios are unlikely to merge within even multiple Hubble times, while mass
ratios near unity merge on nearly a dynamical time.  We note that
Eqn.~\ref{eq:tdf_fit} does indeed tend toward $\tdf \approx \tdn$ for
$\msat=\mhost$ and $\eta \approx 0.5$; it therefore should give a reasonable
approximation even for our untested regime of $\msat/\mhost>0.3$.

We have
covered a range of circularities -- $0.3 \le \eta \le 1.0$ -- that includes the
most likely values based on analyses of orbits in dark matter simulations ($\eta
\approx 0.5$; e.g., \citealt{benson2005, zentner2005a, khochfar2006}).
Equation~(\ref{eq:tdf_fit}) should not be used for $\eta \lesssim 0.2$ because
such orbits pass sufficiently close to the center of the host halo on their
first pericentric passage that interaction with the central galaxy is important
and will decrease the merging timescale.

The range of orbital energies considered here -- $0.65 \le r_c(E)/ \rvir \le
1.0$ -- also covers the peak values of distributions seen in cosmological
simulations.  We have limited ourselves to one specific scaling of dark matter
halo scale radii -- $a \propto M^{0.5}$ -- that is motivated by analyses of
cosmological $N$-body simulations (e.g., \citealt{bullock2001}), which show that
concentrations scale as $c \propto M^{-0.13}$.  \citet{taffoni2003} found that
the dependence of $\tdf$ on the satellite concentration is relatively weak:
using their eq.~27, varying $c_{\rm sat}/c_{\rm host}$ between 1 and 2 results
in a change in $\tdf$ of only $\approx 20\%$.

\subsection{Comparison to Previous Work}

\label{sec:compare}

\subsubsection{Exponent b: dependence on mass ratio}

One important difference between equations~(\ref{eq:dfSAM}) and
(\ref{eq:tdf_fit}) is the non-linear dependence on the mass ratio in
equation~(\ref{eq:tdf_fit}).  Our fitted value of $b=1.3$ reflects an
important difference between point masses and realistic dark matter
satellites sinking in a larger host potential, as the dark matter
halos continually lose mass via tidal stripping and gravitational
shocking against the host potential.  The effective mass of the
satellite over its lifetime is therefore smaller (sometimes
substantially smaller) than its mass when entering the host halo.  It
is therefore not surprising that extended satellites sink more slowly
than the standard formula predicts, leading to $b > 1$.

The exact exponent $b$ turns out to be difficult to model from first
principles as it depends on a number of factors.  As an example,
\citet{fujii2006} and \citet{fellhauer2007} have shown that the
effective mass (with regards to dynamical friction) of a sinking
satellite is actually larger than the instantaneous bound mass because
the mass that has been recently stripped also contributes to the drag
force on the satellite.  We see the same effect in our simulations,
significantly complicating any effort to assume that either the bound
mass or the instantaneous tidal mass is the relevant mass at every
time.

\subsubsection{Parameters c and d: dependence on satellite orbits}
\label{subsubsec:c_and_d}

A common choice for the angular momentum dependence in
equation~(\ref{eq:tdf_fit}) is $\tdf \propto \eta^{0.78}$
\citep{somerville1999, cole2000}, which \citet{lacey1993} found to be a
good match to their integration of the orbit-averaged equations for energy
and angular momentum loss for a point mass subhalo due to dynamical
friction in an isothermal potential.  \citet{van-den-bosch1999} found a
somewhat weaker angular momentum dependence in their numerical simulations
(using point-mass satellites and an isothermal potential), $\tdf \propto
\eta^{0.53}$, while \citet{taffoni2003} provide a more complex fitting
formula for both point-mass and live satellites with realistic internal
structure in an NFW potential.

We find that an exponential dependence -- $\tdf \propto \exp(c \, \eta)$ with
$c=1.9$ -- provides a better fit to our simulations than a pure power law: a fit
in which we force $\tdf \propto \eta^{0.78}$ has a standard deviation of 14 \%
and a maximum deviation of 42\% when compared to the simulations (compared to
6.7 \% and 12 \%, respectively, when using our fit).  It is important to note,
however, that it is the long $\tdf$ for the $\eta=0.78$ and $\eta=1.0$
simulations that result in better agreement using the exponential fit rather
than the power-law fit.  If we restrict our attention to runs with $0.33 \le
\eta \le 0.65$, we find that $\tdf \propto \eta^{0.78}$ does indeed provide a
good fit to our simulation results.

Our revised fitting formula has a somewhat weaker dependence on $\eta$
for $\eta \la 0.5$ and a stronger dependence for $\eta \ga 0.5$
relative to that found by previous work with point-mass satellites.
Both of these limits are plausible: for small $\eta$ (nearly radial
orbits), a slight change in $\eta$ should have a negligible effect on
$\tdf$ since the orbits will have small pericentric distances.  For
large $\eta$ (nearly circular orbits), live satellites still lose mass
at large radii, reducing their mass and lengthening their $\tdf$
relative to an equivalent point-mass satellite.  The same is true for
live satellites on more radial orbits but to a lesser degree, as
angular momentum losses correlate well with pericentric passages (see
Fig.~\ref{fig:ang_all}); the net effect creates a stronger dependence
on $\eta$ for $\eta \ga 0.6$ for live satellites than for point-mass
satellites.

For the orbital energy dependence of the merger timescale, we find
that $d=1.0$ is the best fit to our simulations.  This is to be
contrasted with the canonical value of $d \approx 2.0$ in prior work.
The latter is appropriate for an isothermal host (e.g., eq.~7.26 of
\citealt{binney1987}), or for {\it point mass} satellites sinking on
circular orbits in a static potential, as can be determined by
integrating equation~(\ref{eq:df}).  This yields $\tdf \propto
(r_c(E)/\rvir)^{1.97}$ for an NFW potential \citep{taffoni2003}.  We
find that for point-mass satellites on circular orbits in a host halo
with a Hernquist density profile and scale radius $a$, the energy
dependence of $\tdf$ has an analytic expression,
\begin{equation}
  \label{eq:1}
  \tdf \propto \sqrt{x}\, \left({1 \over 10}\, x^2+{1 \over 3}\,
    x-1+{\tan^{-1} \sqrt{x} \over \sqrt{x}}\right), 
\end{equation}
where $x \equiv r_c(E)/a$.  Equation~(\ref{eq:1}) can be well-approximated by
$\tdf \propto r_c(E)^{2.3}$ for $0.2 \le x \le 30$, comparable to $d=2$.  The
difference between these results for point masses and our best-fit value of
$d=1$ lies in our study of live satellites and in our definition of $\msat$.
Because of ambiguities in defining $\msat$ at small radii when significant tidal
stripping has occurred, we have chosen to use virial quantities exclusively in
equation~(\ref{eq:tdf_fit}), including for the satellite mass $\msat$.  Were we
to use {\it locally}-defined quantities such as $\msat (t)$, we would expect
that $\msat (t) \propto r(t)$ for an isothermal sphere, so that the merger time
evaluated as a function of radius is given by $\tdf \propto r^2(t)/\msat (t)
\propto r(t)$.  It is therefore not surprising that we find $d \sim 1$ to be a
better match than $d \approx 2$ to our numerical results.

\subsubsection{Numerical comparison}

\begin{figure}
  \centering
  \includegraphics[scale=0.55]{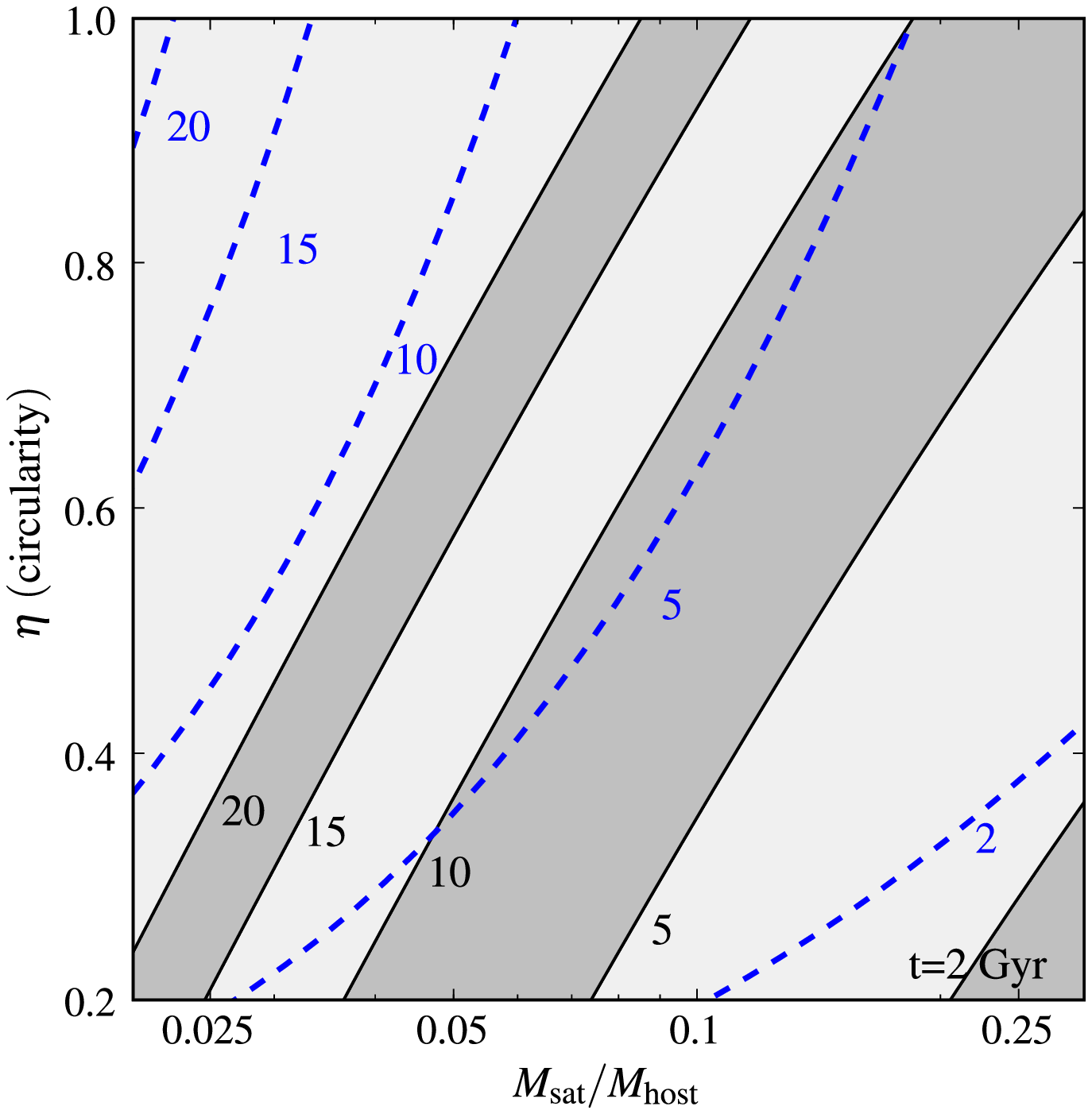}
  \caption{Contours of equal $\tdf$ in the space of satellite-to-host mass ratio
    $\msat/\mhost$ and orbital circularity $\eta$, assuming $r_c(E)=\rvir$.  The
    black solid curves represent our fitting formula eqs.~(\ref{eq:tdf_fit}) and
    (\ref{eq:abcd}); the blue dashed curves correspond to the same SAM shown in
    Fig.~\ref{fig:sim_sam_fit}.  The labelled times are for a host halo with
    $\mhost=10^{12}\, M_\odot$ and $a=40$ kpc.  The SAM generally under-predicts
    $\tdf$, with the largest discrepancy at large $\eta$ and small
    $\msat/\mhost$.}
  \label{fig:tcontour}
\end{figure}

In Figs.~\ref{fig:sim_sam_fit} and \ref{fig:tcontour} we quantify the
differences between equation~(\ref{eq:tdf_fit}) and prior work discussed in the
previous two subsections.  Fig.~\ref{fig:sim_sam_fit} compares $\tdf$(SAM) from
a fiducial SAM using equation~(\ref{eq:dfSAM}) and $\tdf$(fit) from our fits to
equation~(\ref{eq:tdf_fit}) for a range of satellite-to-host mass ratio
$\msat/\mhost$ and orbital circularity ($\eta=0.25$, 0.5, and 0.75 for the
solid, dashed and dot-dashed curves, respectively).  The SAM here assumes
$f_{df}=1$, $\Theta_{\rm orb}=\eta^{0.78}$, and $\logl=\ln(1+\mhost/\msat)$.
The curves illustrate that the fiducial SAM {\it underestimates} the merger
timescale at all masses we have considered regardless of the orbital
circularity.  Moreover, the disagreement grows as the satellite mass decreases
relative to the host: at $\msat/\mhost=0.2$ the SAM timescale is too small by
approximately 50\%, while the discrepancy increases to a factor of 3 at
$\msat/\mhost=0.025$.  If a Coulomb logarithm of $\Lambda=1+(\mhost/\msat)^2$ is
used \citep{somerville1999}, the discrepancy increases further by approximately
a factor of 2 for $\msat \ll \mhost$.  In order to provide a better match to
simulations or observations, some groups adjust the normalization of $\tdf$ by
setting $f_{df} \not= 1$ (e.g., \citealt[$f_{df}=2$]{de-lucia2007};
\citealt[$f_{df}=1.5$]{bower2006}; and \citealt[$f_{df}=0.8$]{nagashima2005}).
Although $f_{df} > 1$ goes in the right direction, Figure \ref{fig:sim_sam_fit}
shows that a constant $f_{df}$ does a poor job of matching the numerical
results.

Fig.~\ref{fig:tcontour} compares the contours of constant $\tdf$ in the
space of mass ratio $\msat/\mhost$ and orbital circularity $\eta$, as
predicted from our fits in equation~(\ref{eq:tdf_fit}) (solid curves) and
the same SAM from Fig.~\ref{fig:sim_sam_fit} (dashed curves).  The two sets
of curves are for the same five merger times: 2, 5, 10, 15, and 20 Gyr.
All orbits that lie to the right of and below a given contour will merge
within the time corresponding to that contour, while those to the left of
and above a contour will not.  Fig.~\ref{fig:tcontour} shows that even the
predictions for orbits with the fastest merging times -- those with low
circularities and large mass ratios (the lower-right portion of the figure)
-- differ between the SAM and our fit.  At larger circularities or smaller
mass ratios, the two predictions deviate significantly.  As an example, the
prediction from our revised formula for 20 Gyr merging times is similar to
the curve for 10 Gyr merging times in the SAM.

Our results can be compared to the fitting formulae provided by
\citet{taffoni2003} [and updated by \citealt{monaco2007}], which are valid
for $0.3 \le r_c(E)/ \rvir \le 0.9$.  The overall trend we find is
qualitatively similar to the results of Taffoni et al. (e.g., their
Fig.~7), but the quantitative details are somewhat different.  We find that
massive satellites ($\msat/\mhost \ga 0.05$) sink more slowly than
predicted by Taffoni et al. while light satellites sink more quickly.  For
example, we predict a timescale that is 25\% shorter than Taffoni et
al. for $\msat/\mhost=0.025$, $r_c(E)/\rvir=0.75$, and $\eta=0.65$ but find
a timescale that is 40\% longer for a satellite with $\msat/\mhost=0.1$ and
identical orbital parameters.  Fig.~7 of Taffoni et al. also shows that the
deviation of live satellite merger times from a fiducial SAM prediction is
not monotonic with $\msat/\mhost$ as is suggested by our
Fig.~\ref{fig:sim_sam_fit} and equation~(\ref{eq:tdf_fit}).  In particular,
for $\msat/\mhost \la 0.001$ and $\msat/\mhost \ga 0.2$, $\tdf$ is
approximately the same in the extended and point mass cases.  We have not
explored mass ratios lower than 1:100, however, because the sinking times
exceed many Hubble times.

After this paper was submitted, \citet[hereafter J07]{jiang2007} independently
investigated the merging timescales of satellites using $N$-body simulations and
came to a similar overall conclusion as that in this paper, namely, that
Eqn.~\ref{eq:dfSAM} generally underpredicts merging timescales.  J07 used a
different $N$-body approach -- cosmological simulations including hydrodynamics
-- and obtained a fitting formula that differs in its details from ours.
Specifically, J07 found $b=1$ (rather than 1.3) and a significantly weaker
dependence on circularity.  The aforementioned differences between the methods
used in this study and J07, as well as differences in definitions (e.g., we
define halo radii relative to $200\,\rho_c$ while J07 use $\Delta_v(z) \,
\rho_c$) make a direct comparison non-trivial.  However, even for our most major
mergers ($\msat/\mhost \ge 0.1$, runs 3x, 5x, and 10x), we still find a
significantly stronger dependence on $\eta$ than J07: whereas the fitting
formula in J07 predicts a difference in $\tdf$ of only 30\% between $\eta=0.46$
and $\eta=1$ orbits, we find a difference of 160\% (runs 5b and 5e).  In
addition, we find that at fixed orbital parameters, the difference in merger
times between our most major merger simulations (the 0.3:1 and 1:5 runs) is
somewhat better fit by $b = 1$ than $b = 1.3$.  Most of the non-linear
dependence on mass-ratio that we find in equation 6 is thus due to the larger
mass ratio (more minor) mergers.  This suggests that the difference between our
preferred value of $b = 1.3$ and J07's value of $b = 1$ may be due to the range
of mass ratios considered.

\section{A Sample Application}

The merger timescale given by equation~(\ref{eq:tdf_fit}) can be applied to
a wide variety of astrophysical problems.  We defer most of these to future
papers.  To give one concrete example, however, we briefly consider the
dissipationless growth of stellar mass for galaxies at the centers of dark
matter halos.

\begin{figure}
  \centering
  \includegraphics[scale=0.55]{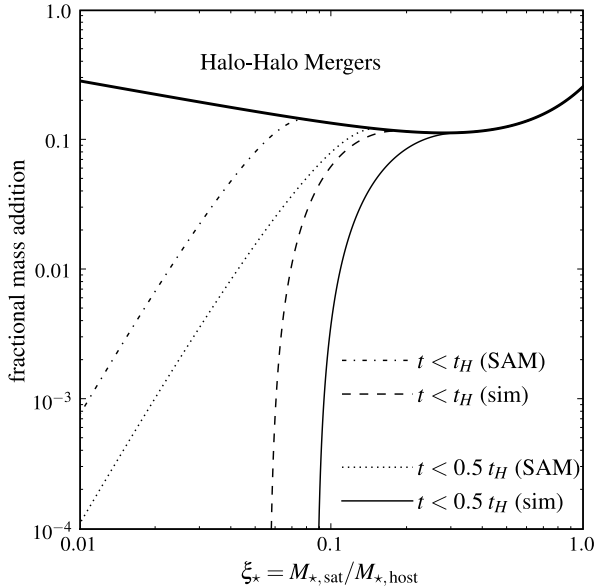}
  \caption{Fraction of stellar mass gained through mergers (i.e.,
    satellite accretion) at the host's virial radius $r_{\rm vir}$
    that makes it down to the central galaxy in time $t$, as a
    function of the satellite-to-host stellar mass ratio
    $\xi_{\star}$.  The thick solid line is the number of halo-halo
    mergers at $r_{\rm vir}$ as a function of mass ratio $\xi_{\star}$
    (normalized to integrate to unity).  We consider the predictions
    of both the fiducial SAM from Fig.~\ref{fig:sim_sam_fit} and our
    eqn.~(\ref{eq:tdf_fit}). The fiducial SAM predicts that
    approximately 40\% more mass is accreted in one Hubble time, all
    of which is in low-$\xi_{\star}$ mergers.}
  \label{fig:dmdlogxi}
\end{figure}

In a finite time period, whether or not galaxies can merge following the
merger of their much more extended host halos depends on $\tdf$ due to
dynamical friction.  Given the dark matter halo-halo merger rate $\dd^2
N_m/\dd\xi\dd z$ at some redshift $z$, defined as the average number of
mergers with mass ratio $\xi=M_2/M_1 \le 1$ leading to halos of mass
$M=M_1+M_2$ per halo per redshift per $\xi$, the corresponding growth of
stellar mass over a time interval $t$ can be computed using
\begin{equation}
  \label{eq:convol}
  {\dd^4 M_{\star} \over \dd \xi \, \dd \eta \, \dd E \, \dd z}= 
  M_{\star}(M_2)  \frac{\dd ^2 N_m}{\dd\xi\, \dd z}
  {\dd ^2 P \over \dd \eta \, \dd E} \, \Theta(t-\tdf) 
\end{equation}
where $\dd^2 P/ \dd \eta \, \dd E$ is the probability of two dark matter halos
merging with orbital circularity $\eta$ and energy $E$ (typically measured when
the two halos' virial radii touch), $\Theta$ is the unit step function, and
$M_{\star}(M_2)$ is the stellar mass as a function of dark matter mass.  The
step function in equation~(\ref{eq:convol}) enforces the fact that only galaxies
in halos with $\tdf \le t$ can merge with the central host galaxy in time $t$. A
full calculation of $\dd M_{\star}/ \dd t$ would require an appropriate integral
of equation~(\ref{eq:convol}) over time $t$, so that all halo-halo mergers are
accounted for.  Results of the full calculation will be presented in
Boylan-Kolchin et al. (in preparation); here we consider instead the simpler
problem of using equation~(\ref{eq:convol}) to calculate the fraction of the
stellar mass accreted at redshift $z$ at $r_{\rm vir}$ that makes it down to the
center within time $t$.  In particular, by integrating
equation~(\ref{eq:convol}), the stellar mass accreted in satellites of different
mass, circularity, or energy can be computed.

We first consider the mass spectrum of merging objects that determines
whether most of the stellar mass added to the galaxy comes from a small
number of massive progenitors or a large number of low-mass progenitors.
This is obtained from equation (\ref{eq:convol}) by integrating over $\eta$
and $E$, taking into account the dependence of $\tdf$ on each of these
orbital parameters.  \citet{zentner2005a} found that $\dd P/\dd \eta
\propto \eta^{1.2} \, (1-\eta)^{1.2}$, which we use in
equation~(\ref{eq:convol}).  For simplicity, we assume that all orbits have
an energy equal to the energy of a circular orbit at the host's virial
radius ($r_c(E)=\rvir$), i.e., $\dd P/ \dd E \, = \,
\delta[E-E_{c}(\rvir)]$.  The halo merger rate is taken from the analysis
of Fakhouri \& Ma (2007, in preparation), who have analyzed merger rates in
the Millennium simulation \citep{springel2005b} and find that
\begin{equation}
  \frac{\dd^2 N_m}{\dd\xi\, \dd z} \propto \xi^{-1.92}\, \exp \left[ \left({\xi
        \over 0.125}\right)^{0.4} \right] \,.
  \label{eq:rmxi}
\end{equation}
Finally, we assume that dark matter halo mass and the stellar mass of its
central galaxy scale as $M_{dm} \propto M_{\star}^{1.5}$, which is appropriate
for massive galaxies in groups and clusters \citep{guzik2002, hoekstra2005,
  mandelbaum2006a}.

Fig.~\ref{fig:dmdlogxi} shows how different dynamical friction
prescriptions lead to different cutoffs at the low mass end for the
fraction of mass assembled via mergers of a given (stellar) mass ratio
$\xi_{\star}$.  We plot both the predictions of our numerical fit in
equation~(\ref{eq:tdf_fit}) and the fiducial SAM from
Fig.~\ref{fig:sim_sam_fit}.  Techincally, we should consider orbits
with $\eta \la 0.2$ separately from those with $\eta \ga 0.2$, as the
dynamics of very radial mergers can be strongly affected by the
stellar components of the merging galaxies at first pericentric
passage, introducing complications that we have not considered in our
current analysis.  For illustrative purposes, however, these orbits,
which account for $\approx 9\%$ of the accreted mass at $\rvir$, are
included in Fig.~\ref{fig:dmdlogxi}.

In all of the calculations in Fig.~\ref{fig:dmdlogxi}, a wide range of
$\xi_{\star} \sim 0.1-1$ contribute to the mass growth of the central
galaxy.  For a canonical SAM dynamical friction formula, however, a
non-negligible amount of extra mass is added in low mass ratio mergers,
relative to the predictions of our formula for $\tdf$.  In fact, $\dd
M_{\star}/ \dd t$ is approximately 40\% larger for $t=t_H$ for the
canonical SAM formula for $\tdf$, and all of the extra mass is accreted in
low-$\xi$ mergers.

\begin{figure}
  \centering
  \includegraphics[scale=0.55]{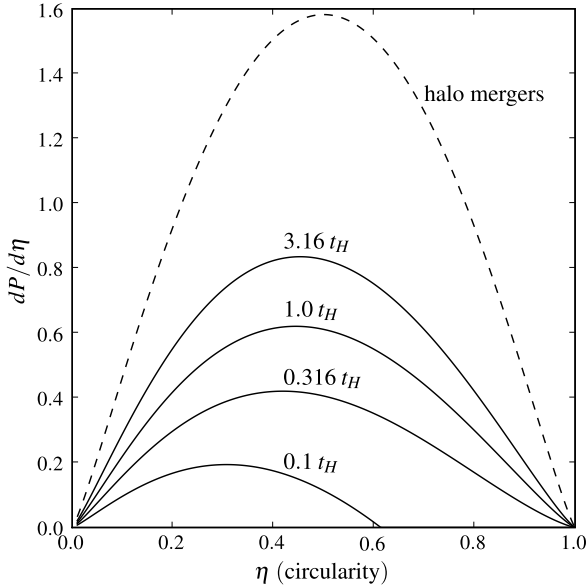}
  \caption{Impact of dynamical friction on the orbital properties of
    satellites that merge with a central galaxy.  The input
    distribution of circularities $\eta$ for halo-halo mergers at
    $r_{\rm vir}$ with $\xi > 10^{-3}$ is shown with the dashed curve
    ($dP/d\eta$, taken from \citealt{zentner2005a}).  The solid curves
    show the predicted distribution of $\eta$ for satellite galaxies
    that merge onto the central galaxy within a specific fraction of a
    Hubble time using equation~(\ref{eq:tdf_fit}).  Higher mass ratio
    and lower angular momentum orbits preferentially accrete, leading
    to a shift in the $dP/d\eta$ distribution to lower circularities.}
  \label{fig:dmdeta}
\end{figure}

Dynamical friction also distorts the orbital distribution of satellites at
$\rvir$ of the host, preferentially selecting low angular momentum orbits
for accretion onto the central galaxy.  This effect is shown in
Fig.~\ref{fig:dmdeta}, which shows the initial distribution of orbital
circularities (dashed curve; taken from Zentner et al. and assumed to be
independent of $\xi$) along with the distribution of circularities for
satellites merging within a given fraction of a Hubble time ($\tdf/t_H=0.1,
\, 0.316,\, 1.0, \, {\rm and} \, 3.16$).  For $\tdf/t_H \gg 1$, the stellar
mass accreted as a function of circularity should approach the circularity
distribution at $\rvir$, as all satellites will eventually merge.  As
$\tdf/t_H$ decreases, however, the peak of the distribution shifts to
smaller and smaller circularities because low angular momentum satellites
merge more rapidly.  Fig.~\ref{fig:dmdeta} shows that the peak of the input
distribution of circularities ($\dd P/\dd \eta$) is at $\eta=0.5$ but that
most of the mass comes in via satellites on orbits with $\eta=0.42$ at
$\tdf/t_H=1$ and with $\eta=0.30$ for $\tdf/t_H=0.1$.  Shifting the peak of
the circularity distribution from 0.5 to 0.42 corresponds to a $30 \%$
reduction in the pericentric distance of the orbit, which can have a
significant effect on the properties of the merger remnant (see, e.g.,
\citealt{boylan-kolchin2006}). 

\section{Conclusions}

Applications of dynamical friction to problems in galaxy formation are often
based on classic results derived for point masses sinking in a galaxy with an
isothermal density profile (e.g., eq.~[\ref{eq:dfSAM}]).  Using direct numerical
simulations, we have shown that live dark matter satellites with realistic
internal structure and mass ratios in the range relevant for galaxy formation
($0.02 \la \msat/\mhost \la 0.3$) take much longer to merge than the
corresponding point-mass satellite results, which are typically used in
semi-analytic galaxy formation models (e.g.,
Figs.~\ref{fig:dftimes_pred}-\ref{fig:tcontour}).  This difference is primarily
because the satellites undergo significant mass loss as they sink deeper in the
potential well of the host halo.  Including a stellar bulge in both the
satellite and host changes the merger time by $\la$ 10\% for typical orbital
circularities (see Fig.~\ref{fig:stars}).  We find that the surprisingly simple
fitting formula given in equations~(\ref{eq:tdf_fit}) and (\ref{eq:abcd})
provides an accurate ($\la 10 \%$ error) fit to the numerically determined
galaxy merging times over the range of mass ratios, orbital circularities, and
orbital energies we have considered.  The dependence of $\tdf$ on mass ratio,
energy, and angular momentum implied by this fit is all somewhat different from
standard assumptions in the literature (see \S \ref{sec:compare}).  We have
chosen to test our results for a range of parameters that are cosmologically
interesting, e.g. for satellite-to-host mass ratios for which the satellite
undergoes significant orbital evolution within a Hubble time, and for orbital
parameters that are relvant according to cosmological dark matter simulations.

The results of this paper should be relevant to a wide range of problems in
galaxy formation and evolution.  For example, halo-halo merger rates as a
function of mass ratio $\msat/\mhost$ from extended Press-Schechter theory or
cosmological simulations show that a significant amount of dark matter mass is
accumulated via accretion of lower-mass halos with $0.02 \la \msat/\mhost \la
0.3$ (e.g., eq.~[\ref{eq:rmxi}] \& Fig.~\ref{fig:dmdlogxi}; \citealt{lacey1993};
\citealt{zentner2007}; Fakhouri \& Ma 2007, in prep).  Understanding the
evolution of the galaxy population in such halos thus requires an accurate
understanding of the effects of dynamical friction in precisely the mass range
where the point mass approximation is inadequate.  To give a few concrete
examples, we have shown that standard merging timescales in the literature
overestimate the growth of stellar mass by satellite accretion by $\approx 40
\%$, with the extra mass gained in low mass ratio mergers
(Fig.~\ref{fig:dmdlogxi}).  In addition, we have quantified the tendency for
satellites that accrete onto a central galaxy to have lower angular momentum
than average: the peak in the circularity distribution shifts from $\approx 0.5$
to $\approx 0.42$ when one considers only satellites that merge onto a central
galaxy within a Hubble time (Fig.~\ref{fig:dmdeta}).

Several limitations of our calculations should be noted.  For orbits with very
low circularity ($\eta \la 0.2$), our results are not applicable because such
orbits pass sufficiently close to the center of the host halo on their first
pericentric passage that interaction with the central galaxy is important and
will decrease the merging timescale.  All but two of our 28 simulations neglect
the effect of the central galaxy in a given dark matter halo on the merging
timescale.  As Fig.~\ref{fig:stars} shows, this is a reasonable approximation
when considering merging from the virial radius (as we have done), but would be
less applicable for studying the detailed evolution of satellites at small radii
in a dark matter halo (e.g. \citealt{ostriker1977}).  Finally, our results do
not take into account drag on ambient gas, which may be important in cluster and
group environments (e.g., \citealt{ostriker1999}).

\vspace{2cm}

We thank the referee for useful suggestions and Volker Springel for making {\sc
  gadget-2} publicly available.  This work used resources from NERSC, which is
supported by the US Department of Energy.  C-PM is supported in part by NSF
grant AST 0407351.  EQ is supported in part by NASA grant NNG05GO22H and the
David and Lucile Packard Foundation.

\bibliographystyle{apj}
\bibliography{df}
\end{document}